\documentclass[twocolumn]{aastex62}
\pdfoutput=1

\usepackage{amsmath}
\usepackage{cancel}
\usepackage{float}

\shorttitle{Updated Opacities for DSEP}
\shortauthors{Boudreaux et al.}

\begin{document}

\title{Correlations between Ca II H\&K Emission and the Gaia M dwarf Gap}

\correspondingauthor{Emily M. Boudreaux}
\email{emily.m.boudreaux.gr@dartmouth.edu,\\emily@boudreauxmail.com}

\author[0000-0002-2600-7513]{Emily M. Boudreaux}
\affiliation{Department of Physics and Astronomy, Dartmouth College, Hanover, NH 03755, USA}

\author[0000-0001-9828-3229]{Aylin Garcia Soto}
\affiliation{Department of Physics and Astronomy, Dartmouth College, Hanover, NH 03755, USA}

\author[0000-0003-3096-4161]{Brian C. Chaboyer}
\affiliation{Department of Physics and Astronomy, Dartmouth College, Hanover, NH 03755, USA}

\received{02/07/2024}
\revised{02/20/2024}
\accepted{02/22/2024}
\submitjournal{ApJ}

\begin{abstract}
  The Gaia M dwarf gap, also known as the Jao Gap, is a novel feature
  discovered in the Gaia DR2 G vs. BP-RP color magnitude diagram. This gap
  represents a 17 percent decrease in stellar density in a thin magnitude band
  around the convective transition mass ($\sim 0.35 M_{\odot}$) on the main
  sequence. Previous work has demonstrated a paucity of Hydrogen Alpha emission
  coincident with the G magnitude of the Jao Gap in the solar neighborhood. The
  exact mechanism which results in this paucity is as of yet unknown; however,
  the authors of the originating paper suggest that it may be the result of
  complex variations to a star's magnetic topology driven by the Jao Gap's
  characteristic formation and breakdown of stars' radiative transition zones.
  We present a follow up investigating another widely used magnetic activity
  metric, Calcium II H\&K emission. Ca II H\&K activity appears to share a
  similar anomalous behavior as H$\alpha$ does near the Jao Gap magnitude. We
  observe an increase in star-to-star variation of magnetic activity near the
  Jao Gap. We present a toy model of a stars magnetic field evolution
  which demonstrates that this increase may be due to stochastic disruptions
  to the magnetic field originating from the periodic mixing events
  characteristic of the convective kissing instabilities which drive the
  formation of the Jao Gap.
\end{abstract}


\section{INTRODUCTION}\label{sec:intro}
The initial mass requirements of molecular clouds collapsing to form stars
results in a strong bias towards lower masses and later spectral classes
during star formation. Partly as a result of this bias and partly as a result
of their extremely long main-sequence lifetimes, M Dwarfs make up approximately
70 percent of all stars in the galaxy \citep{Winters2019}. Moreover, many
planet search campaigns have focused on M Dwarfs due to the relative ease of
detecting small planets in their habitable zones \citep[e.g.][]{Nut08}. M
Dwarfs then represent both a key component of the galactic stellar population
as well as the most numerous possible set of stars which may host habitable
exoplanets. Given this key location M Dwarfs occupy in modern astronomy it is
important to have a thorough understanding of their structure and evolution.

\citet{Jao2018} discovered a novel feature in the Gaia Data Release 2 (DR2)
$G_{BP}-G_{RP}$ color-magnitude-diagram. Around $M_{G}=10$ there is an
approximately 17 percent decrease in stellar density of the sample of stars
\citet{Jao2018} considered. Subsequently, this has become known as either the
Jao Gap, or Gaia M Dwarf Gap. Following the initial detection of the Gap in DR2
the Gap has also potentially been observed in 2MASS \citep{Skrutskie2006,
Jao2018}; however, the significance of this detection is quite weak and it
relies on the prior of the Gap's location from Gaia data. The Gap is
also present in Gaia Early Data Release 3 (EDR3) \citep{Jao2021}. These EDR3
and 2MASS data sets then indicate that this feature is not a bias inherent to
DR2.

The Gap is generally attributed to convective instabilities in the cores of
stars straddling the fully convective transition mass (0.3 - 0.35 M$_{\odot}$)
known as convective kissing instabilities \citep{vanSaders2012, Baraffe2018}.
These instabilities interrupt the normal, slow, main sequence luminosity
evolution of a star and result in luminosities lower than expected from the
main sequence mass-luminosity relation \citep{Jao2020}.

The Jao Gap, inherently a feature of M Dwarf populations, provides an enticing
and unique view into the interior physics of these stars \citep{Feiden2021}.
This is especially important as, unlike more massive stars, M Dwarf seismology
is infeasible due to the short periods and extremely small
magnitudes which both radial and low-order low-degree non-radial seismic waves
are predicted to have in such low mass stars \citep{Rodriguez-Lopez2019}. The
Jao Gap therefore provides one of the only current methods to probe the
interior physics of M Dwarfs.

The magnetic activity of M dwarfs is of particular interest due to the
theorised links between habitability and the magnetic environment which a
planet resides within \citep[e.g.][]{Lammer2012,Gallet2017, Kislyakova2017}. M
dwarfs are known to be more magnetically active than earlier type stars
\citep{Saar1985,Astudillo-Defru2017,Wright2018} while simultaneously this same
high activity calls into question the canonical magnetic dynamo believed to
drive the magnetic field of solar-like stars (the $\alpha\Omega$ dynamo)
\citep{Shulyak2015}. One primary challenge which M dwarfs pose is that stars
less than approximately 0.35 M$_{\odot}$ are composed of a single convective
region. This denies any dynamo model differential rotation between adjacent
levels within the star. Alternative dynamo models have been proposed, such as
the $\alpha^{2}$ dynamo along with modifications to the $\alpha\Omega$ dynamo
which may be predictive of M dwarf magnetic fields \citep{Chabrier2006,
Kochukhov2021, Kleeorin2023}.

Despite this work, very few studies have dived specifically into the magnetic
field of M dwarfs at or near the convective transition region . This is not
surprising as that only spans approximately a 0.2 magnitude region
in the Gaia BP-RP color magnitude diagram and is therefore populated by a
relatively small sample of stars. 

\citet{Jao2023} identify the Jao Gap as a strong discontinuity point for
magnetic activity in M dwarfs. Two primary observations from their work are
that the Gap serves as a boundary where very few active stars, in their sample
of 640 M dwarfs, exist below the Gap and that the overall downward trend of
activity moving to fainter magnitudes is anomalously high in within the 0.2 mag
range of the Gap. \citeauthor{Jao2023} Figures 3 and 13 make this paucity in
H$\alpha$ emission particularly clear. Based on previous work from
\citet{Spada2020, Curtis2020, Dungee2022} the authors propose that the
mechanism resulting in the reduced fraction of active stars within the Gap is
that as the radiative zone dissipates due to core expansion, angular momentum
from the outer convective zone is dumped into the core resulting in a faster
spin down than would otherwise be possible. Effectively the core of the star
acts as a sink, reducing the amount of angular momentum which needs to be lost
by magnetic breaking for the outer convective region to reach the same angular
velocity. Given that H$\alpha$ emission is strongly coupled magnetic activity
in the upper chromosphere \citep{Newton2016, Kumar2023} and that a star's angular velocity
is a primary factor in its magnetic activity, a faster spin down will serve to
more quickly dampen H$\alpha$ activity.

In addition to H$\alpha$ the Calcium Fraunhaufer lines may be used to trace the
magnetic activity of a star. These lines originate from magnetic heating of the
lower chromosphere driven by magnetic shear stresses within the star.
Both \citet{Perdelwitz2021} and \citet{Boudreaux2022} present calcium emission
measurements for stars spanning the Jao Gap. In this paper we search for
similar trends in the Ca II H\& K emission as \citeauthor{Jao2023} see in the
H$\alpha$ emission. In Section \ref{sec:results} we investigate the empirical
star-to-star variability in emission and quantify if this could be due to noise
or sample bias; in Section \ref{sec:modeling} we present a simplified toy model
which shows that the mixing events characteristic of convective kissing
instabilities could lead to increased star-to-star variability in activity as
is seen empirically.

%


\section{Correlation}\label{sec:results}
Using Ca II H\&K emission data from \citet{Perdelwitz2021} and
\citet{Boudreaux2022} \citep[quantified using the $R'_{HK}$
metric][]{Middelkoop1982, Rutten1984} we investigate the correlation between
the Jao Gap magnitude and stellar magnetic activity. We are more statistically
limited here than past authors have been due to the requirement for high
resolution spectroscopic data when measuring Calcium emission.

\begin{figure}
  \centering
  \includegraphics[width=0.45\textwidth]{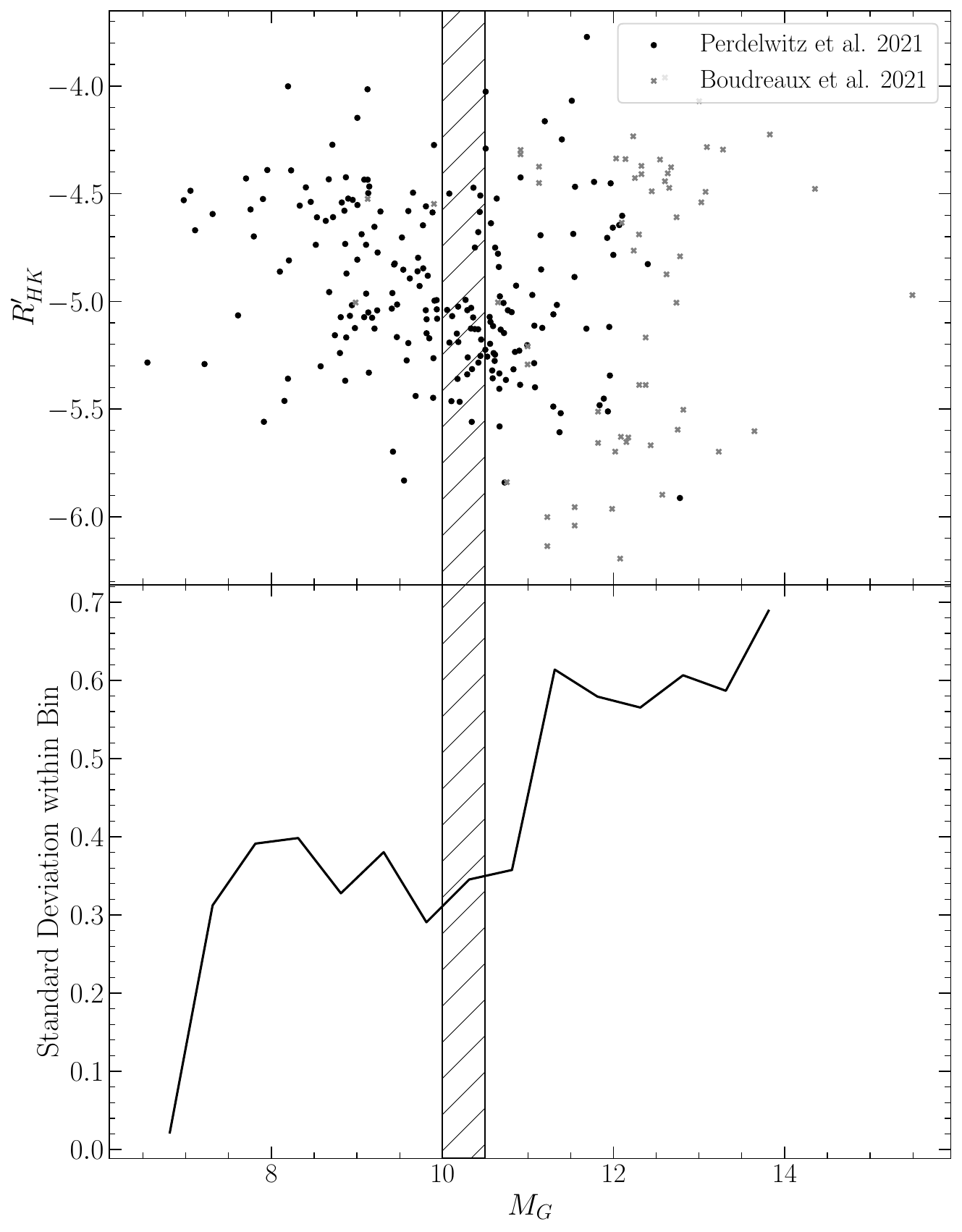}
  \caption{Merged Dataset from \citet{Perdelwitz2021, Boudreaux2022}. Note the
  increase in the spread of $R'_{HK}$ around the Jao Gap Magnitude (top).
  Standard deviation of Calcium emission data within each bin. Note the
  discontinuity near the Jao Gap Magnitude (bottom). The location of the Gap
  as identified in literature is shown by the hatched region ($\sim$ 10-10.5 $M_{G}$). Potential
  explanations for the disagreement in magnitude are discussed in detail
  in Section \ref{sec:coincident}.}
  \label{fig:initData}
\end{figure}

The merged dataset is presented in Figure \ref{fig:initData}. The
sample overlap between \citet{Perdelwitz2021} and \citet{Boudreaux2022} is
small (only consisting of five targets). For those five targets there is an
approximately 1.5 percent average difference between measured $\log(R'_{HK})$
values, with measurements from \citeauthor{Boudreaux2022} biased to be slightly
more negative than those from \citeauthor{Perdelwitz2021}

There is a visual discontinuity in the spread of stellar activity
below the Jao Gap magnitude. Further discussion of why there may be
disagreement between the observed magnitude of the gap and the discontinuity
which we identify may be found in Section \ref{sec:coincident}. In order to
quantify the significance of this discontinuity we measure the false alarm
probability of the change in standard deviation.

\begin{figure}
  \centering
  \includegraphics[width=0.45\textwidth]{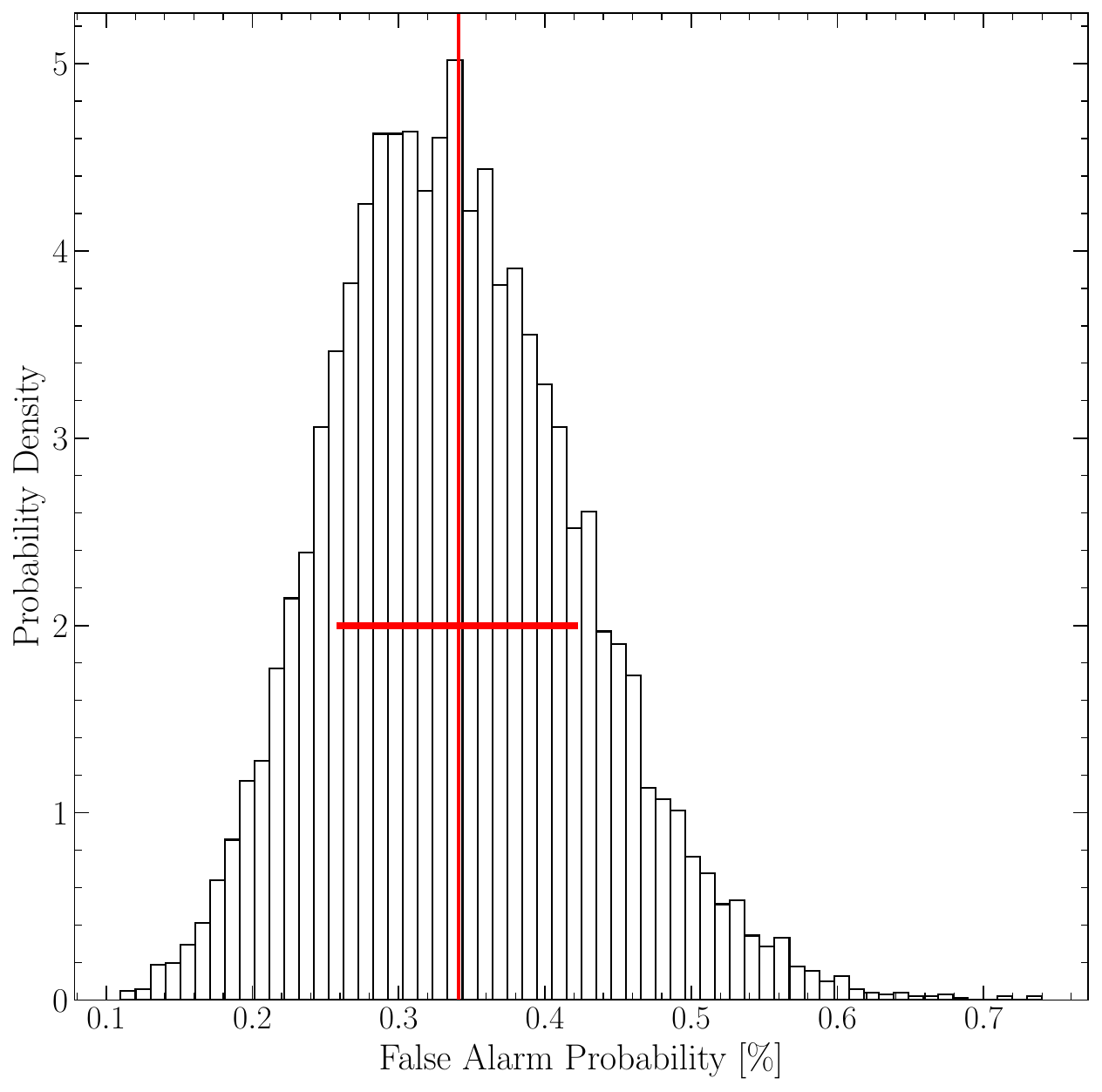}
  \caption{Probability distribution of the false alarm probability for the
  discontinuity seen in Figure \ref{fig:initData}. The mean of this
  distribution is $0.341\%\pm^{0.08}_{0.08}$.}
  \label{fig:dist}
\end{figure}

First we split the merged dataset into bins with a width of 0.5 mag. In each bin we
measure the standard deviation about the mean of the data. The results of this
are shown in Figure \ref{fig:initData} (bottom). In order to measure the false alarm
probability of this discontinuity we first resample the merged calcium
emission data based on the associated uncertainties for each datum as
presented in their respective publications. Then, for each of these ``resample
trials'' we measure the probability that a change in the standard deviation of
the size seen would happen purely due to noise. Results of this test are show in
in Figure \ref{fig:dist}. 

This rapid increase star-to-star variability would only arise due purely to
noise $0.3\pm0.08$ percent of the time and is therefore likely either a true
effect or an alias of some sample bias.

\begin{figure}
  \centering
  \includegraphics[width=0.45\textwidth]{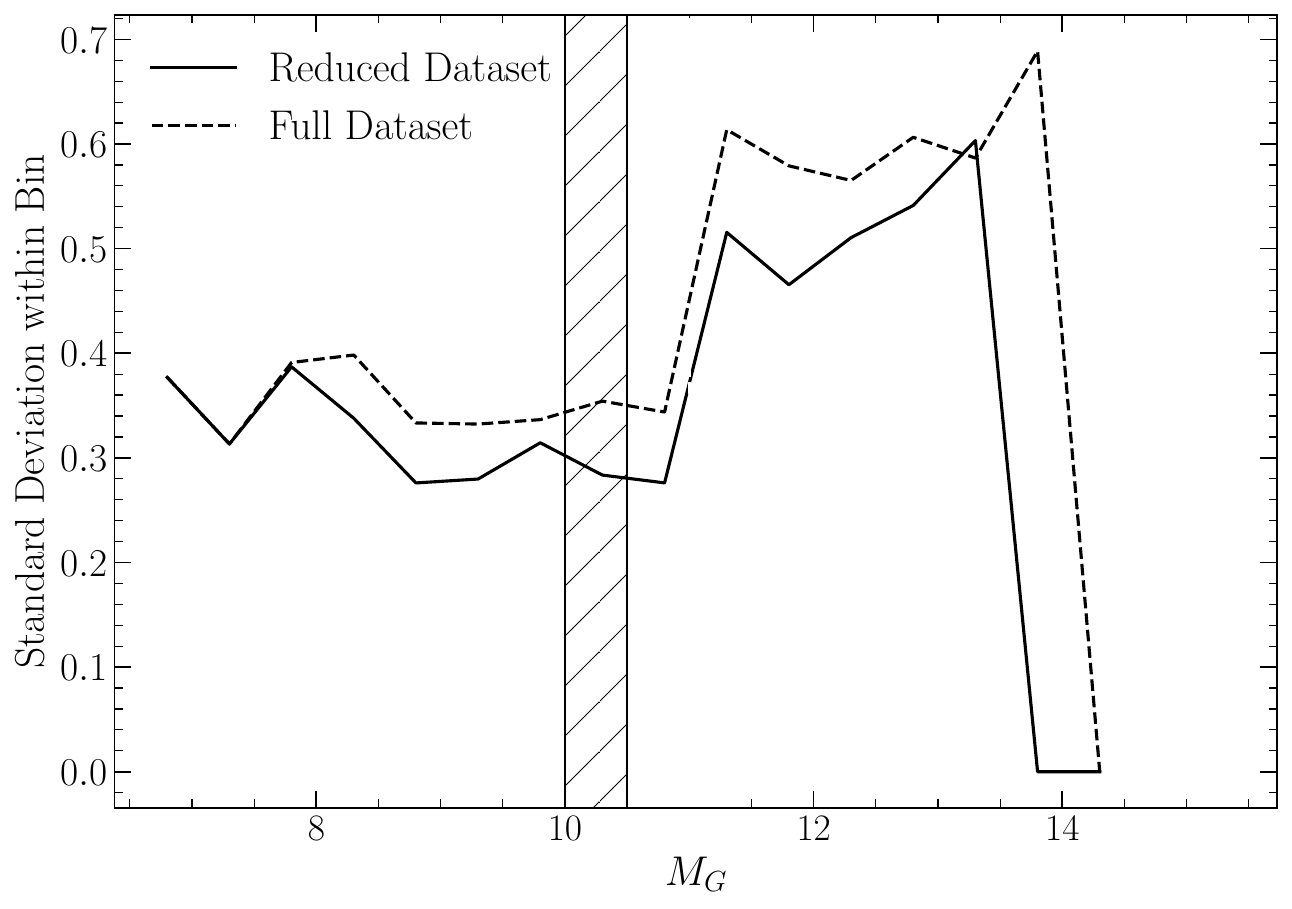}
  \caption{Spread in the magnetic activity metric for the merged sample with
  any stars $\log(R'_{HK}) > -4.436$ filtered out.The location of the Gap
  as identified in literature is shown by the hatched region ($\sim$ 10-10.5 $M_{G}$).}
  \label{fig:reduced}
\end{figure}

If the observed increase in variability is not due to a sample bias and rather
is a physically driven effect then there is an obvious similarity between these
findings and those of \citet{Jao2023}. Specifically we find a increase in
variability below the magnitude of the Gap. Moreover, this variability
increase is primarily driven by an increase in the number of low activity stars
(as opposed to an increase in the number of high activity stars). We can
further investigate the observed change in variability for only low activity
stars by filtering out those stars at or above the saturated threshold for
magnetic activity. \citet{Boudreaux2022} identify $\log(R'_{HK}) = -4.436$ as
the saturation threshold. We adopt this value and filter out all stars where
$\log(R'_{HK}) \geq -4.436$. Applying the same analysis to this reduced dataset
as was done to the full dataset we still find a discontinuity at the same
location (Figure \ref{fig:reduced}). This discontinuity is of a smaller
magnitude and consequently is more likely to be due purely to noise, with a
$7\pm0.2$ percent false alarm probability. This false alarm probability is
however only concerned with the first point after the jump in variability. If
we consider the false alarm probability of the entire high variability region
then the probability that the high variability region is due purely to noise
drops to $1.4\pm0.04$ percent.

Further, various authors have shown that the strength of Calcium II
H\&K emission may evolve over month to year timescales
\citep[e.g.][]{Rauscher2006, Perdelwitz2021, Cretignier2024}. Targets from
\citet{Boudreaux2022} were observed an average of only four times and over year
long timescales. Therefore, the nominal $\log(R'_{HK})$ values derived in that
work may be biased by stellar variability. However, the scale of observed
variabilty in the activity metric is signifigantly smaller than the
star-to-star activity variability addressed here and therfore activity cycles
are not expected to be of particular relevance. Specifically, the amplitude of
variability is generally $\Delta \log(R'_{HK}) \lessapprox 0.2$ wherase in this
work we address variability on the order of $\Delta \log(R'_{HK}) \lessapprox 2$.

We observe a strong, likely statistically significant, discontinuity in the
star-to-star variability of Ca II H\&K emission below the magnitude
of the Jao Gap. However, modeling is required to determine if this discontinuity
may be due to the same underlying physics.

\subsection{Conicidence with the Jao Gap Magnitude}\label{sec:coincident}
\begin{figure}
  \centering
  \includegraphics[width=0.45\textwidth]{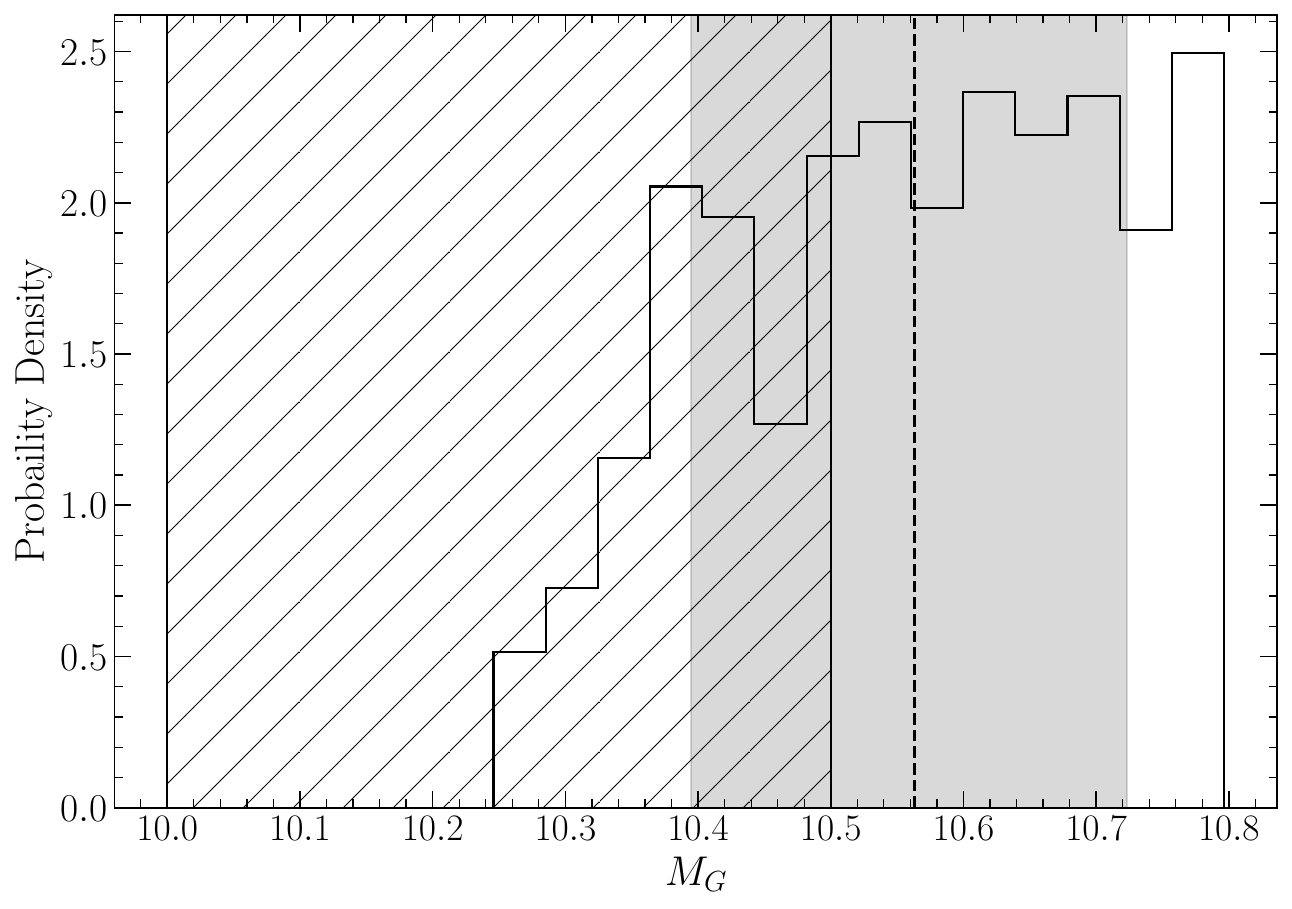}
  \caption{Probability density distribution of discontinuity location as
  identified in the merged dataset. The dashed line represents the mean of the
  distribution while the shaded region runs from the 16th percentile to the
  84th percentile of the distribution. This distribution was built from 10,000
  independent samples where the discontinuity was identified as the highest
  value in the gradient of the standard deviation.The location of the Gap
  as identified in literature is shown by the hatched region ($\sim$ 10-10.5 $M_{G}$).}
  \label{fig:GapLocationMC}
\end{figure}

While the observed increase in variability seen here does not seem to be
coincident with the Jao Gap --- instead appearing to be approximately 0.5 mag
fainter, in agreement with what is observed in \citet{Jao2023} --- a number of
complicating factors prevent us from falsifying that the these two features are
not coincident. \citeauthor{Jao2023} find, similar to the results presented
here, that the paucity of $H\alpha$ emission originates below the Gap.
Moreover, we use a 0.5 magnitude bin size when measuring the star-to-star
variability which injects error into the positioning of any feature in
magnitude space. We can quantify the degree of uncertainty the magnitude bin
choice injects by conducting Monte Carlo trials where bins are randomly shifted
redder or bluer. We conduct 10,000 trials where each trial involves sampling a
random shift to the bin start location from a normal distribution with a
standard deviation of 1 magnitude. For each trial we identify the discontinuity
location as the maximum value of the gradient of the standard deviation (this
is the derivative of the data in Figures \ref{fig:deviation} \&
\ref{fig:reduced}). Some trials result in the maximal value lying at the 0th
index of the magnitude array due to edge effects, these trials are rejected
(and account for 11\% of the trials). The uncertainty in the identified
magnitude of the discontinuity due to the selected start point of the magnitude
bins reveals a $1\sigma = \pm$0.32 magnitude uncertainty in the location of the
discontinuity (Figure \ref{fig:GapLocationMC}). Finally, all previous studies
of the M dwarf Gap \citep{Jao2018, Jao2021, Mansfield2021, Boudreaux2022,
Jao2023} demonstrate that the Gap has a color dependency, shifting to fainter
magnitudes as the population reddens and consequently an exact magnitude range
is ill-defined. Therefore we cannot falsify the model that the discontinuity in
star-to-star activity variability is coincident with the Jao Gap magnitude.

\subsection{Rotation}
It is well known that star's magnetic activity tend to be correlated with their
rotational velocity \citep{Vaughan1981, Newton2016, Astudillo-Defru2017,
Houdebine2017, Boudreaux2022}; therefore, we investigate whether there is a
similar correlation between Gap location and rotational period in our dataset.
All targets from \citet{Boudreaux2022} already have published rotational
periods; however, targets from \citet{Perdelwitz2021} do not necessarily have
published periods. Therefore, we derive photometric rotational periods for
these targets here. Given the inherent heterogeneity of M Dwarf stellar
surfaces \citep{Boisse2011, Robertson2020} we are able to determine the
rotational period of a star through the analysis of active regions. Various
methodologies can be employed for this purpose, including the examination of
photometry and light curves \citep[e.g.,][]{Newton2016}, and the observation of
temporal changes in the strength of chromospheric emission lines such as Ca II
H \& K or H$\alpha$ \citep[e.g.,][]{2019A&A...623A..24F,2023MNRAS.518.3147K}.
In this work, new rotational periods are derived from TESS 2-minute cadence
data\footnote{Some M Dwarfs lacking a documented rotational period did not have
sufficient TESS data to yield fiducial rotational periods}.

Due to both the large frequency and amplitudes of M dwarf flaring rates the
photometric period can prove difficult to measure --- as frequency directly
correlates with periodicity. Thus, following the process described in
\citet{2023AJ....165..192G}, we utilize two methods in this paper to reduce the
effect of flares. One method uses \texttt{stella} a python package which
implements a series of pre-trained convolutional neural networks (CNNs) to
remove flare-shaped features in a light curve \citep{FeinsteinStella2020}. The
second method separates a star's photometry into 10 minute bins to account for
misshapen flares which \texttt{stella} is known to be biassed against detecting.

\texttt{stella}  employs a diverse library of models trained with varying
initial seeds \citep{FeinsteinFlare2020,FeinsteinStella2020}. The Convolutional
Neural Networks in \texttt{stella} are trained on labeled TESS 2-min for both
flares and non-flares. For the purposes of this paper, we use an ensemble of
100 models in \texttt{stella}'s library to optimize model performance
\citep[][for further detail]{FeinsteinFlare2020}. \texttt{stella} scores
flairs with a probability of between 0 to 1 --- where higher values indicate a
higher confidence that a feature is a flare. Here we adopt a score of 0.5 as
the cutoff threshold, all features with a score of 0.5 or greater are classed
as flares and removed \citep[e.g.][]{FeinsteinFlare2020}.

Furthermore, we also bin the data from a 2-min to 10-min cadence using the
python package \texttt{lightkurve}'s binning function
\citep{LightkurveCollaborationLightkurve2018,GeertBarentsenKeplerGO2020}. This
further reduces any flaring-contribution that might have been missed by
\texttt{stella}\footnote{This is relevant for flares that are misshapen at the
start or break in the dataset due to missing either the ingress or egress.}.
Subsequently, we filter photometry, only retaining data whos residuals are
less than 4 times the root-mean-square deviation.  

Gaussian processes for modeling the periods are based on
\citet{AngusInferring2018} for the subset of M Dwarfs with no fiducial periods.
The \texttt{starspot} \ package is adapted for light curve analysis
\citep{AngusRuthAngus2021, Angus2023}. Our Gaussian process kernel function
incorporates two stochastically-driven simple harmonic oscillators,
representing primary ($P_\textrm{rot}$) and secondary ($P_\textrm{rot}/2$)
rotation modes. First, we implement the Lomb-Scargle periodogram within
\texttt{starspot} to initially estimate the period. After which, we create a
maximum a posteriori (MAP) fit using \texttt{starspot} to generate a model for
stellar rotation. To obtain the posterior of the stellar rotation model, we use
Markov Chain Monte Carlo (MCMC) sampling using the \texttt{pymc3} package
\citep{SalvatierProbabilistic2016} within our adapted \texttt{starspot}
version. All rotational periods are presented in Table \ref{tab:dataset}. Our
final sample contains 187 stars with measured rotational periods. We derive new
rotational periods for 7 of these. 

\begin{deluxetable*}{ccccccccc}
  \tablehead{\colhead{ID} & \colhead{G Mag} & \colhead{V Mag} & \colhead{K Mag} & \colhead{$\log(R'_{HK})$} & \colhead{e\_Log(R'$_{HK}$)} & \colhead{Ro} & \colhead{prot} & \colhead{r\_prot}\\ \colhead{ } & \colhead{$\mathrm{mag}$} & \colhead{$\mathrm{mag}$} & \colhead{$\mathrm{mag}$} & \colhead{ } & \colhead{ } & \colhead{ } & \colhead{$\mathrm{d}$} & \colhead{ }}
\startdata
2MASS J00094508-4201396 & 12.14 & 13.659 & 8.223 & -4.339 & 0.001 & 0.009 & 0.859 & Bou22 \\
2MASS J00310412-7201061 & 12.301 & 13.648 & 8.445 & -5.388 & 0.003 & 0.928 & 80.969 & Bou22 \\
2MASS J01040695-6522272 & 12.447 & 13.95 & 8.532 & -4.489 & 0.001 & 0.006 & 0.624 & Bou22 \\
2MASS J02004725-1021209 & 12.778 & 14.113 & 9.092 & -4.791 & 0.001 & 0.188 & 14.793 & Bou22 \\
2MASS J02014384-1017295 & 13.026 & 14.477 & 9.189 & -4.54 & 0.001 & 0.034 & 3.152 & Bou22 \\
2MASS J02125458+0000167 & 12.096 & 13.58 & 8.168 & -4.635 & 0.001 & 0.048 & 4.732 & Bou22 \\
2MASS J02411510-0432177 & 12.251 & 13.79 & 8.246 & -4.427 & 0.001 & 0.004 & 0.4 & Bou22 \\
2MASS J03100305-2341308 & 12.23 & 13.5 & 8.567 & -4.234 & 0.001 & 0.028 & 2.083 & Bou22 \\
2MASS J03205178-6351524 & 12.087 & 13.433 & 8.195 & -5.629 & 0.004 & 1.029 & 91.622 & Bou22 \\
2MASS J05015746-0656459 & 10.649 & 12.196 & 6.736 & -5.005 & 0.002 & 0.875 & 88.5 & Bou22
\enddata
\caption{First 10 rows of the dataset used in this work. This data is avalible as a machine readable supliment to this article.}
\label{tab:dataset}

\end{deluxetable*}

\begin{figure}
  \centering
  \includegraphics[width=0.45\textwidth]{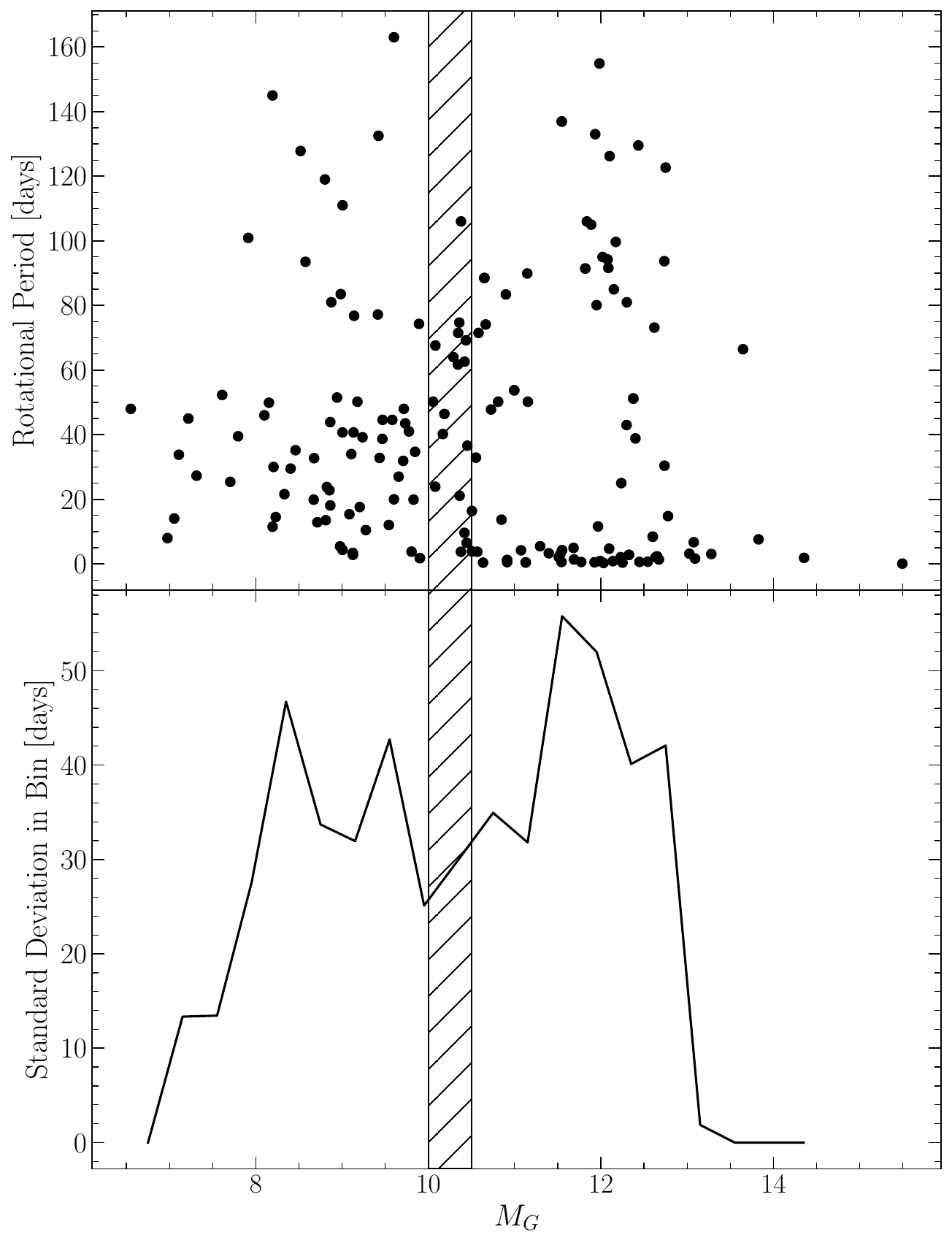}
  \caption{Rotational Periods against G magnitude for all stars with rotational
  periods (top). Standard deviation of rotational period within magnitude bin (bottom).The location of the Gap
  as identified in literature is shown by the hatched region ($\sim$ 10-10.5 $M_{G}$).}
  \label{fig:rotationalSignifigance}
\end{figure}

One might expect a decrease in mean rotational period around the magnitude of
the Gap, due to the slight decrease in magnetic activity. However, there is no
statistically significant correlation between rotational period and G
magnitude which we can detect given our sample size (Figure
\ref{fig:rotationalSignifigance}). Rotational period is however, not the ideal
parametrization to use, as magnetic activity is more directly related to the
Rossby number ($Ro$). Using the empirical calibration presented in
\citet{Wright2018} (Equation \ref{eqn:tauc}) we find the mixing timescale for
each star such that the Rossby Number is defined as $Ro = P_{rot}/\tau_{c}$.

\begin{equation}\label{eqn:tauc}
  \tau_{c} = 0.64 + 0.25 * (V-K)
\end{equation}

\begin{figure}
  \centering
  \includegraphics[width=0.45\textwidth]{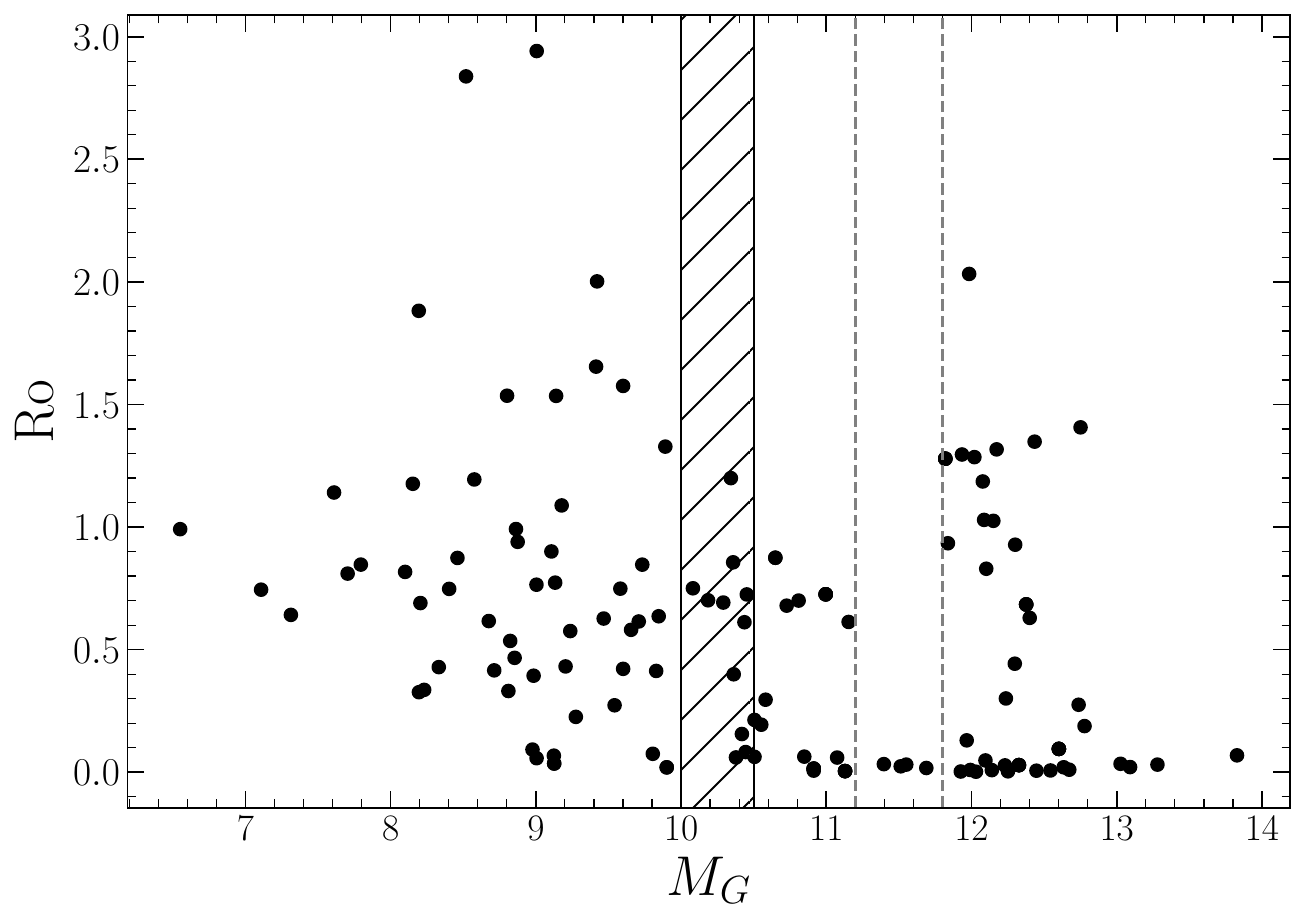}
  \caption{Rossby number vs. G magnitude for all stars with rotational periods
  and V-K colors on Simbad. Dashed lines represent the hypothesized region of decreased rotation.The location of the Gap
  as identified in literature is shown by the hatched region ($\sim$ 10-10.5 $M_{G}$).}
  \label{fig:rossby}
\end{figure}

When we compare Rossby number to G magnitude (Figure \ref{fig:rossby}) we find
that there may be a slight paucity of rotation coincident with the decrease in
spread of the activity metric. We quantify the statistical significance of this
drop by building a Gaussian kernel density estimator (kde) based on the data
outside of this range, and then resampling that kde 10000 times for each data
point in the theorized paucity range. The false alarm probability that that drop
is due to noise is then the product of the fraction of samples which are less
than or equal to the value of each data point. We find that there is a 0.022
percent probability that this dip is due purely to noise.

\subsection{Limitations}
There are two primary limitation of our dataset. First, we only have 264 star
in our dataset (with measured $R'_{HK}$, 187 with rotational periods) limiting
the statistical power of our analysis. This is primarily due to the relative
difficulty of obtaining Ca II H\&K measurements compared to obtaining $H\alpha$
measurements. Reliable measurements require both high spectral resolutions (R
$\sim$ 16000) and a comparatively blue wavelength range \footnote{wrt. to what
many spectrographs cover. There is no unified resource listing currently
commissioned spectrographs; however, it is somewhat hard to source glass which
transmits well at H\&K wavelengths limiting the lower wavelength of most
spectrographs.}.

Additionally, the sample we do have does not extend to as low mass as would be
ideal. This presents a degeneracy between two potential causes for the observed
increased star-to-star variability. One option, as presented above and
elaborated on in the following section, is that this is due to kissing
instabilities. However, another possibility is that this increased variability
is intrinsic to the magnetic fields of fully convective stars. This alternate
option may be further supported by the shape of the magnetic activity spread vs.
G magnitude relation. Convective kissing instabilities are not expected to
continue to much lower masses than the fully convective transition mass. The
fact that the increase in variance which we observe continues to much fainter
magnitudes would therefore be somewhat surprising in a purely convective kissing
instability driven framework (though the degeneracy between potentially
physically driven increase in variance and increase in variance due to the
noise-magnitude relation complicates attempts to constrain this.) There is
limited discussion in the literature of overall magnetic field strength
spanning the fully convective transition mass; however, \citet{Shulyak2019}
present estimated magnetic field strengths for 47 M dwarfs, spanning a larger
area around the convective transition region and their dataset does not
indicate a inherently increased variability for fully convective stars.

\section{Modeling}\label{sec:modeling}
One of the most pressing questions related to this work is whether or not the
increased star-to-star variability in the activity metric and the Jao Gap,
which are coincident in magnitude, are driven by the same underlying mechanism.
The challenge when addressing this question arises from current computational
limitations. Specifically, the kinds of three dimensional
magneto-hydrodynamical simulations --- which would be needed to derive the
effects of convective kissing instabilities on the magnetic field of the star
--- are infeasible to run over gigayear timescales while maintaining thermal
timescale resolutions needed to resolve periodic mixing events.

In order to address this and answer the specific question of \textit{could
kissing instabilities result in increased star-to-star variability of the
magnetic field}, we adopt a very simple toy model. Kissing instabilities result
in a transient radiative zone separating the core of a star (convective) from its
envelope (convective). When this radiative zone breaks down two important
things happen: one, the entire star becomes mechanically coupled, and two,
convective currents can now move over the entire radius of the star.
\citet{Jao2023} propose that this mechanical coupling may allow the star's core
to act as an angular momentum sink thus accelerating a stars spin down and
resulting in anomalously low H$\alpha$ emission. 

Regardless of the exact mechanism by which the magnetic field may be affected,
it is reasonable to expect that both the mechanical coupling and the change to
the scale of convective currents will have some effect on the star's magnetic
field. On a microscopic scale both of these will change how packets of charge
within a star move and may serve to disrupt a stable dynamo. Therefore, in the
model we present here we make only one primary assumption: \textit{every mixing
event may modify the star's magnetic field by some amount}. Within our model
this assumption manifests as a random linear perturbation applied to some base
magnetic field at every mixing event. The strength of this perturbation is 
sampled from a normal distribution with some standard deviation, $\sigma_{B}$.

\begin{figure*}
  \centering
  \includegraphics[width=0.9\textwidth]{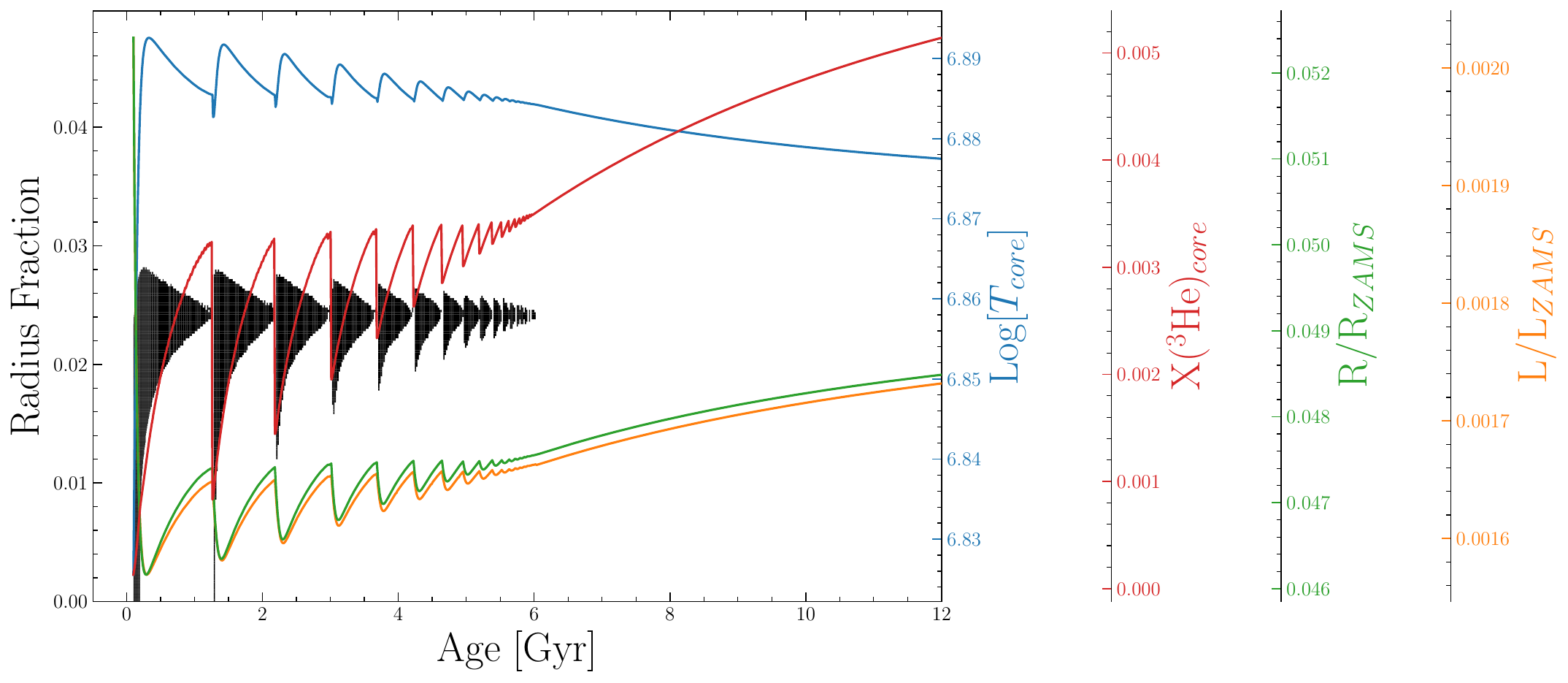}
  \caption{Kippenhan-Iben diagram for a 0.345 solar mass star. Note the
  periodic mixing events (where the plotted curves peak).}
  \label{fig:kippenhan}
\end{figure*}

Synthetic stars are sampled from a grid of stellar models evolved using the
Dartmouth Stellar Evolution Program (DSEP) with similar parameters to those
used in \citet{Boudreaux2023}. Each stellar model was evolved using a high
temporal resolution (timesteps no larger than 10,000 years) and typical
numerical tolerances of one part in $10^5$. Each model was based on a GS98
\citep{Grevesse1998} solar composition with a mass range from 0.3 M$_{\odot}$
to 0.4 M$_{\odot}$. Finally, models adopt OPLIB high temperature radiative
opacities, Ferguson 2004 low temperature radiative opacities, and include both
atomic diffusion and gravitational settling. A Kippenhan-Iben diagram showing
the structural evolution of a model within the Gap is shown in Figure
\ref{fig:kippenhan}.

\begin{figure}
  \centering
  \includegraphics[width=0.45\textwidth]{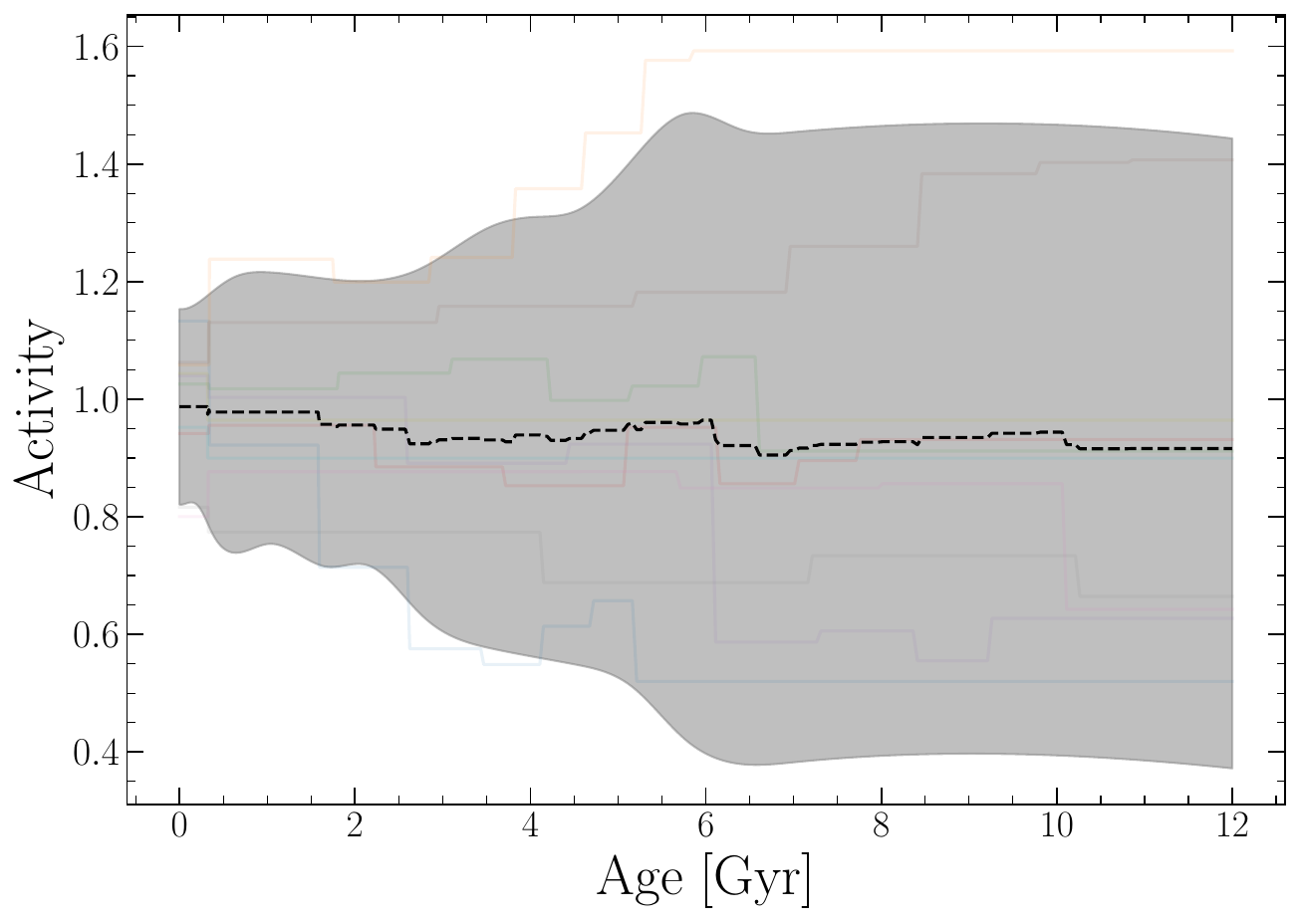}
  \caption{Example of the toy model presented here resulting in increased
  divergence between stars magnetic fields. The shaded region represents the
  maximum spread in the two point correlation function at each age.}
  \label{fig:simpleB}
\end{figure}

Each synthetic star is assigned some base magnetic activity ($B_{0} \sim
\mathcal{N}(1, \sigma_{B})$) and then the number of mixing events before some age $t$
are counted based on local maxima in the core temperature. The toy magnetic
activity at age $t$ for the model is given in Equation \ref{eqn:activity}. An
example of the magnetic evolution resulting from this model is given in Figure
\ref{fig:simpleB}. Fundamentally, this model presents magnetic
activity variation due to mixing events as a random walk and therefore results will
increasing divergence over time.

\begin{align}\label{eqn:activity}
  B(t) = B_{0} + \sum_{i}B_{i} \sim \mathcal{N}(1, \sigma_{B}) 
\end{align}

\begin{figure}
  \centering
  \includegraphics[width=0.45\textwidth]{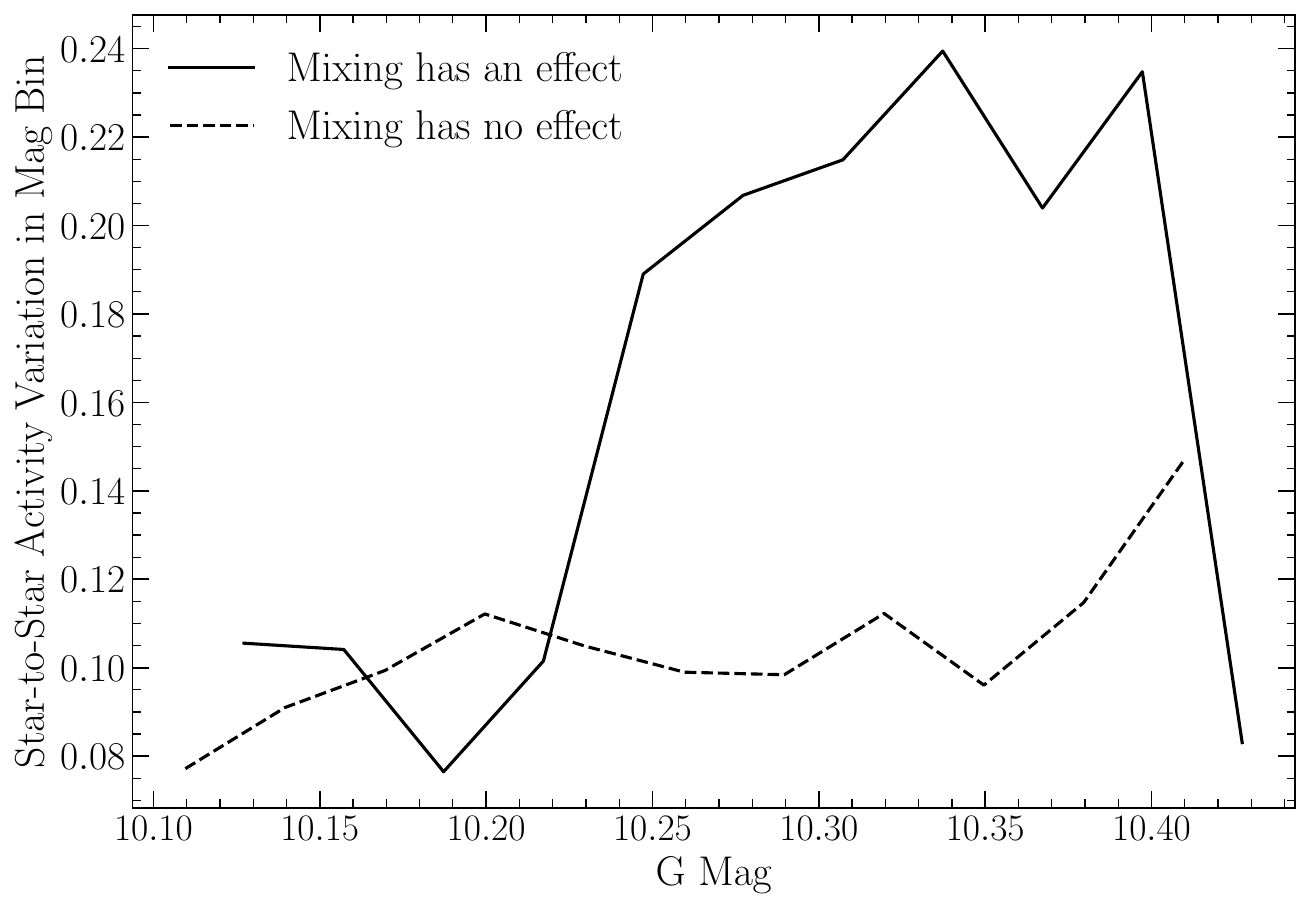}
  \caption{Toy model results showing a qualitatively similar discontinuity in the star-to-star magnetic activity variability.}
  \label{fig:model}
\end{figure}

Applying the same analysis to these models as was done to the observations as
described in Section \ref{sec:results} we find that this simple model results
in a qualitatively similar trend in the standard deviation vs. Magnitude graph
(Figure \ref{fig:model}). In order to reproduce the approximately 50 percent
change to the spread of the activity metric observed in the combined dataset in
section \ref{sec:results} a distribution with a standard deviation of 0.1 is
required when sampling the change in the magnetic activity metric at each
mixing event. This corresponds to 68 percent of mixing events modifying the
activity strength by 10 percent or less. The interpretation here is important:
what this qualitative similarity demonstrates is that it may be reasonable to
expect kissing instabilities to result in the observed increased star-to-star
variation. Importantly, we are not able to claim that kissing instabilities
\textit{do} lead to these increased variations, only that they reasonably
could. Further modeling, observational, and theoretical efforts will be needed
to more definitively answer this question.

\subsection{Limitations}
The model presented in this paper is very limited and it is important to keep
these limitations in mind when interpreting the results presented here. Some of
the main challenges which should be leveled at this model are the assumption
that the magnetic field will be altered by some small random perturbation at
every mixing event. This assumption was informed by the large number of free
parameters available to a physical star during the establishment of a large
scale magnetic field and the associated likely stochastic nature of that
process. However, it is similarly believable that the magnetic field will tend
to alter in a uniform manner at each mixing event. For example, since
differential rotation is generally proportional to the temperature gradient
within a star and activity is strongly coupled to differential rotation then it
may be that as the radiative zone reforms over thermal timescales the
homogenization of angular momentum throughout the star results in overall lower
amounts of differential rotation each after mixing event than would otherwise
be present.

Moreover, this model does not consider how other degenerate sources of magnetic
evolution such as stellar spin down, relaxation, or coronal heating may effect
star-to-star variability. These could conceivably lead to a similar increase in
star-to-star variability which is coincident with the Jao Gap magnitude as the
switch from fully to partially convective may effect efficiency of these
process.

Additionally, there are challenges with this toy model that originate from the
stellar evolutionary model. Observations of the Jao Gap show that the feature
is not perpendicular to the magnitude axis; rather, it is inversely
proportional to the color. No models of the Jao Gap published at the time of
writing capture this color dependency and \textit{what causes this color
dependency} remains one of the most pressing questions relating to the
underlying physics. This non captured physics is one potential explanation for
why the magnitude where our model predicts the increase in variability is not
in agreement with where the variability jump exists in the data.

Finally, we have not considered detailed descriptions of the dynamos of stars.
The magnetohydrodynamical modeling which would be required to model the
evolution of the magnetic field of these stars at thermal timescale resolutions
over gigayears is currently beyond the ability of practical computing.
Therefore future work should focus on limited modeling which may inform the
evolution of the magnetic field directly around the time of a mixing event.

\section{Conclusion}\label{sec:conclusion}
It is, at this point, well established that the Jao Gap may provide a unique view of the interiors of stars for which other probes, such as seismology, fail. However, it has only recently become clear that the Gap may lend insight into not just structural changes within a star but also into the magnetic environment of the star.
\citet{Jao2023} presented evidence that the physics driving the Gap might additionally result in a paucity of H$\alpha$ emission. These authors propose potential physical mechanisms which could explain this paucity, including the core of the star acting as an angular momentum sink during mixing events.

Here we have expanded upon this work by probing the degree and variability of
Calcium II H\&K emission around the Jao Gap. We lack the same statistical
power of \citeauthor{Jao2023}'s sample; however, by focusing on the
star-to-star variability within magnitude bins we are able to retain
statistical power. We find that there is an anomalous increase in variability
at a G magnitude of $\sim 11$. This is only slightly below the observed mean gap magnitude.

Additionally, we propose a simple model to explain this variability. Making the
assumption that the periodic convective mixing events will have some small but
random effect on the overall magnetic field strength we are able to
qualitatively reproduce the increase activity spread in a synthetic population
of stars.

\acknowledgments{
  This work has made use of the NASA astrophysical data system (ADS). We would
  like to thank Elisabeth Newton, Aaron Dotter, and Gregory Feiden for their
  support and for useful discussion related to the topic of this paper.
  Additionally, we would like to thank Keighley Rockcliffe, Kara Fagerstrom,
  and Isabel Halstead for their useful discussion related to this work.
  Finally, we would like to thank the referee for their careful reading and
  critique of this article. We acknowledge the support of a NASA grant (No.
  80NSSC18K0634). 
}
\acknowledgments
\acknowledgments


\software{
  The Dartmouth Stellar Evolution Program (\texttt{DSEP}) \citep{Dotter2008},
	\texttt{BeautifulSoup} \citep{richardson2007beautiful},
	\texttt{mechanize} \citep{chandra2015python},
	\texttt{FreeEOS} \citep{Irwin2012},
	\texttt{pyTOPSScrape} \citep{Boudreaux22},
  \texttt{lightkurve} \citep{LightkurveCollaborationLightkurve2018},
  \texttt{stella} \citep{FeinsteinStella2020},
  \texttt{starspot} \citep{AngusRuthAngus2021,Angus2023}
}

\bibliography{ms}{}
\bibliographystyle{aasjournal}

\end{document}